\documentclass[twocolumn]{aastex631}

\begin{document}

\title{First X-ray and radio polarimetry of the neutron star low-mass X-ray binary GX~17$+$2}

\author{Unnati Kashyap}
\affiliation{Department of Physics and Astronomy, Texas Tech University, Lubbock, TX 79409-1051, USA} 

\author{Thomas J. Maccarone}
\affiliation{Department of Physics and Astronomy, Texas Tech University, Lubbock, TX 79409-1051, USA}

\author{Eliot C. Pattie}
\affiliation{Department of Physics and Astronomy, Texas Tech University, Lubbock, TX 79409-1051, USA}

\author{Mason Ng}
\affiliation{Department of Physics, McGill University, 3600 rue University, Montr\'{e}al, QC H3A 2T8, Canada}
\affiliation{Trottier Space Institute, McGill University, 3550 rue University, Montr\'{e}al, QC H3A 2A7, Canada}

\author{Swati Ravi}
\affiliation{MIT Kavli Institute for Astrophysics and Space Research, Massachusetts Institute of Technology, Cambridge, MA 02139, USA}

\author{Alexandra J. Tetarenko}
\affiliation{Department of Physics and Astronomy, University of Lethbridge, Lethbridge, Alberta, T1K 3M4, Canada}

\author{Pau Bosch Cabot}
\affiliation{Department of Physics and Astronomy, University of Lethbridge, Lethbridge, Alberta, T1K 3M4, Canada}

\author{Herman L. Marshall}
\affiliation{MIT Kavli Institute for Astrophysics and Space Research, Massachusetts Institute of Technology, Cambridge, MA 02139, USA}



\begin{abstract}
We report the first polarimetric results of the neutron star (NS) low-mass X-ray binary (LMXB) Z-source GX~17$+$2 using the Imaging X-ray Polarimetry Explorer (IXPE) and the Very Large Array (VLA). We find that the X-ray source was polarized at  ${\rm PD} = 1.9\pm 0.3 \%$ ($1\sigma$ errors) with a polarization angle of ${\rm PA} = 11\pm4 \degr$ ($1\sigma$ errors). Simultaneous Nuclear Spectroscopic Telescope Array (NuSTAR) observations show that the source was in the normal branch (NB) during our IXPE observations. The X-ray spectro-polarimetry results suggest a source geometry comprising an accretion disk component, a Comptonization component, and a reflection component. The VLA radio polarization study shows a ${\rm PD} = 2.2\pm 0.2 \%$ with a Faraday-corrected intrinsic polarization angle of $1\pm 5 \degr$, which is an indication of the jet axis. Thus, we find the estimated X-ray PA from the source is consistent with the radio PA. We discuss the accretion geometry of the Z-source in light of our X-ray spectro-polarimetry and radio findings.

\end{abstract}



\section{Introduction} \label{sec:intro}
A weakly magnetized neutron star low-mass X-ray binary (NS LMXB) accretes matter from a low-mass companion star ($M<1\,M_\odot$) via Roche-lobe overflow \citep{2023hxga.book..120B}. Based on the shape of the path that they trace out in the X-ray color-color diagram (CCD) or the hardness-intensity diagrams (HID), they are historically divided into two classes, Z and atoll \citep{1989A&A...225...79H}. The primary difference between the Z and atoll classes is the source luminosity. The highly luminous Z-sources accrete at a higher rate close to the Eddington limit relative to the low-accreting atoll sources \citep{1994ApJS...92..511V,2014MNRAS.443.3270M}. The atoll CCDs are characterized by the spectrally harder island and spectrally softer banana states. Z-sources, on the other hand, trace out a Z-shaped pattern in their CCDs, with the upper, middle, and lower branches in a Z pattern referred to as the horizontal branch (HB), normal branch (NB), and flaring branch (FB), respectively \citep{1989A&A...225...79H}. These Z-sources are further classified into Cyg-like and Sco-like sources (see e.g. \cite{2010A&A...512A...9B,2012A&A...546A..35C}): The Cyg-like sources (e.g., Cyg~X-2, GX~5$-$1, and GX~340$+$0) exhibit the complete Z-tracks with prominent Z branches. On the other hand, Sco-like sources (e.g., Sco~X-1, GX~17$+$2, and GX~349$+$2) are scarcely seen in the HB. The Sco-like sources exhibit a prominent flaring behavior (in the FB), unlike the Cyg-like sources, which exhibit weak flaring \citep{2012MmSAI..83..170C}. Furthermore, sources switching between atoll and Z-like behavior have also been reported  (see e.g. \cite{2009AAS...21360303L,2010ApJ...719..201H})

The previous Imaging X-ray Polarimetry Explorer (IXPE) polarization studies showed that the observed polarization in the case of  Z-sources is attributed to the accretion disk, the boundary layer (BL)/spreading layer (SL) \citep{2022MNRAS.514.2561G, 2024A&A...684A..62F, 2025A&A...696A.181B},  and the reflection off the disk atmosphere \citep{1993MNRAS.260..663M,1996MNRAS.283..892P} or wind \citep{2025A&A...694A.230N}. A correlation between the polarization properties and the source spectral state in both persistent and transient Z sources, such as GX 5$-$1 \citep{2024AA...684A.137F}, GX~340$+$0 \citep{2024arXiv240519324B}, and XTE J1701$-$462 \citep{2023A&A...674L..10C}, has also been observed, where sources in the HB exhibit higher PD ($\sim4-5\%$) compared to sources in the NB ($\lesssim2\%$).

\subsection{Z-source type jets}

Z-source type NS LMXBs are observed to produce persistent radio emission in all branches, attributed to a persistent radio jet. This jet luminosity is observed to correlate with their spectral branches, where the flaring branch hosts the faintest jets, and the horizontal branch the brightest \citep[e.g.,][]{Migliari2007ApJ...671..706M}. The persistent jets are believed to be similar to the compact steady jets observed in low accretion rate/hard state X-ray binaries, based in part on their apparent optical thickness up to millimeter wavelengths \citep[e.g.,][]{DiazTrigo2021A&A...650A..37D}. In addition to the compact jets, discrete ejecta can also be launched by Z-sources \citep{Fomalont2001ApJ...558..283F,Spencer2013}, though the pattern of ejecta events in these sources is not well understood. 

\subsection{The Z-source type NS LMXB GX~17$+$2}

 The Z-source type NS LMXB GX~17$+$2, with an estimated spin period $\sim$ 3.4 ms of the central NS \citep{1997ApJ...490L.157W}, is known to have a low-inclination \citep[25 - 38\degr;][]{2017ApJ...836..140L} estimated from X-ray reflection modeling however, no confirmation of this from optical or infrared data exists for this source.  Previous studies report that the continuum spectrum of GX~17$+$2 is well modeled by a single temperature blackbody and a multi-color disk blackbody component along with a weak Comptonized component \citep{2012ApJ...756...34L}. Detections of reflection features have also been reported from the source \citep{2010ApJ...720..205C,2012ApJ...756...34L,2017ApJ...836..140L}. 
A previous BeppoSAX study of GX~17$+$2 also reported the presence of  a hard tail in the spectrum at energies above $\sim$ 30 keV in the
HB, which disappears as the source transitions towards the NB \citep{2000ApJ...544L.119D}. Additionally, the source is known to exhibit aperiodic variabilities during all the branches of its Z track \citep{2025JHEAp..45..250T}. \\

In this paper, we report the first IXPE observations of the Z-source type NS LMXB GX~17$+$2 performed from 2024 Oct 23 to 2024 Oct 26. The source was observed simultaneously with NICER and NuSTAR. The Karl G. Jansky Very Large Array (VLA) observed GX~17$+$2 on 2024 Aug 08. 
In \S\ref{sec:obs}, we outline the details of the observations and the data analysis methods. In \S\ref{sec:results}, we present the results obtained from the
spectro-polarimetric analyses. We discuss our results in \S\ref{sec:discussion} and summarize them in \S\ref{sec:summary}.

\section{Observations and data reductions} 
\label{sec:obs}
\begin{figure}
\centering
\includegraphics[width=0.5\textwidth]{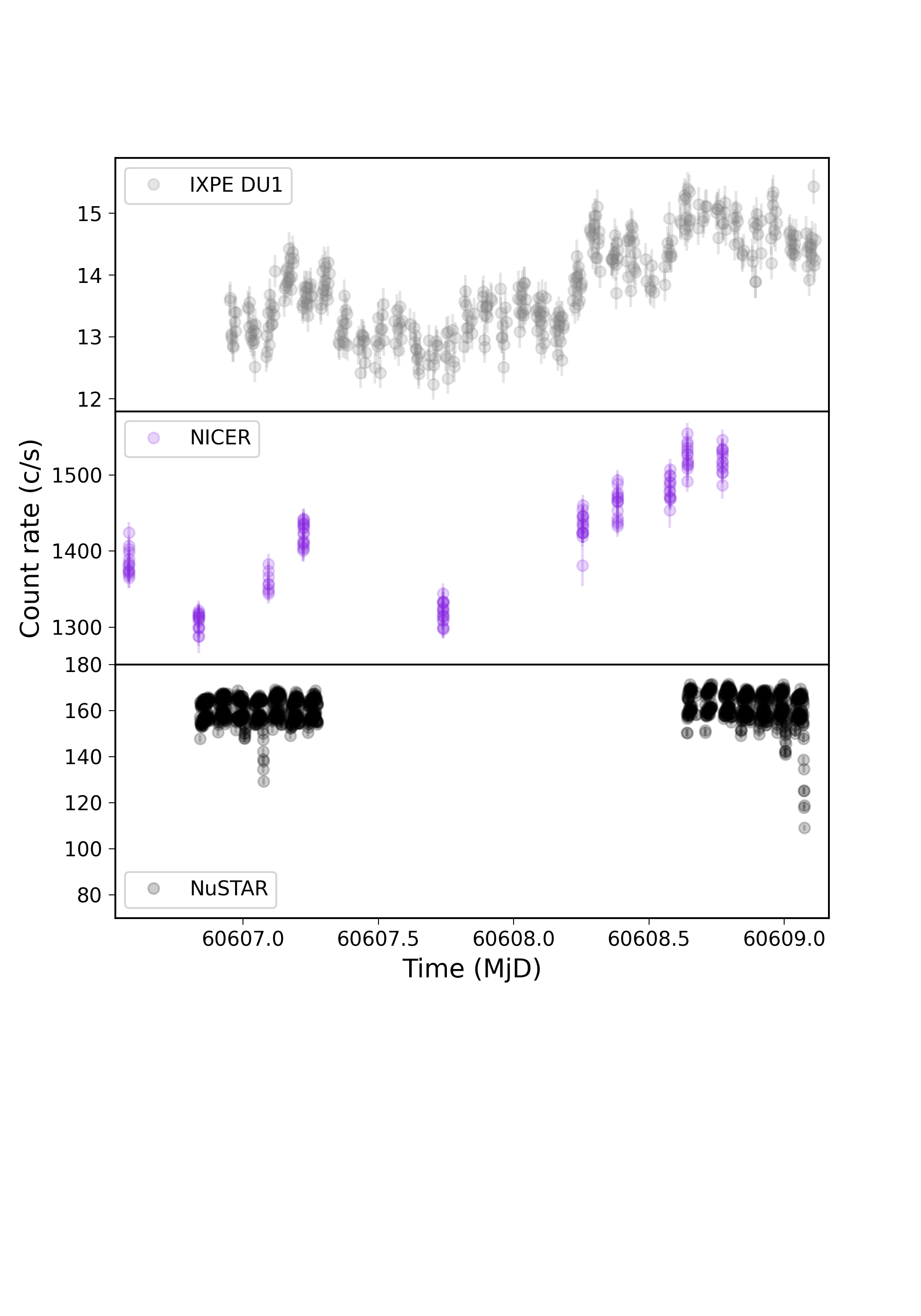}
\caption{First panel: IXPE (2--8~keV) light curve of GX~17$+$2. Time bins of 128 s are used. Second Panel: NICER (0.5--10~keV) light curve of GX~17$+$2. Time bins of 8 s are used. Third Panel: NuSTAR (3.0--79.0~keV) light curve during observation 1 and observation 2 of GX~17$+$2. Time bins of 60 s is used. The source was observed by NICER and NuSTAR simultaneously with the IXPE observations.
}
\label{fig:lc}
\end{figure}

\begin{figure}
\centering
\includegraphics[width=0.5\textwidth]{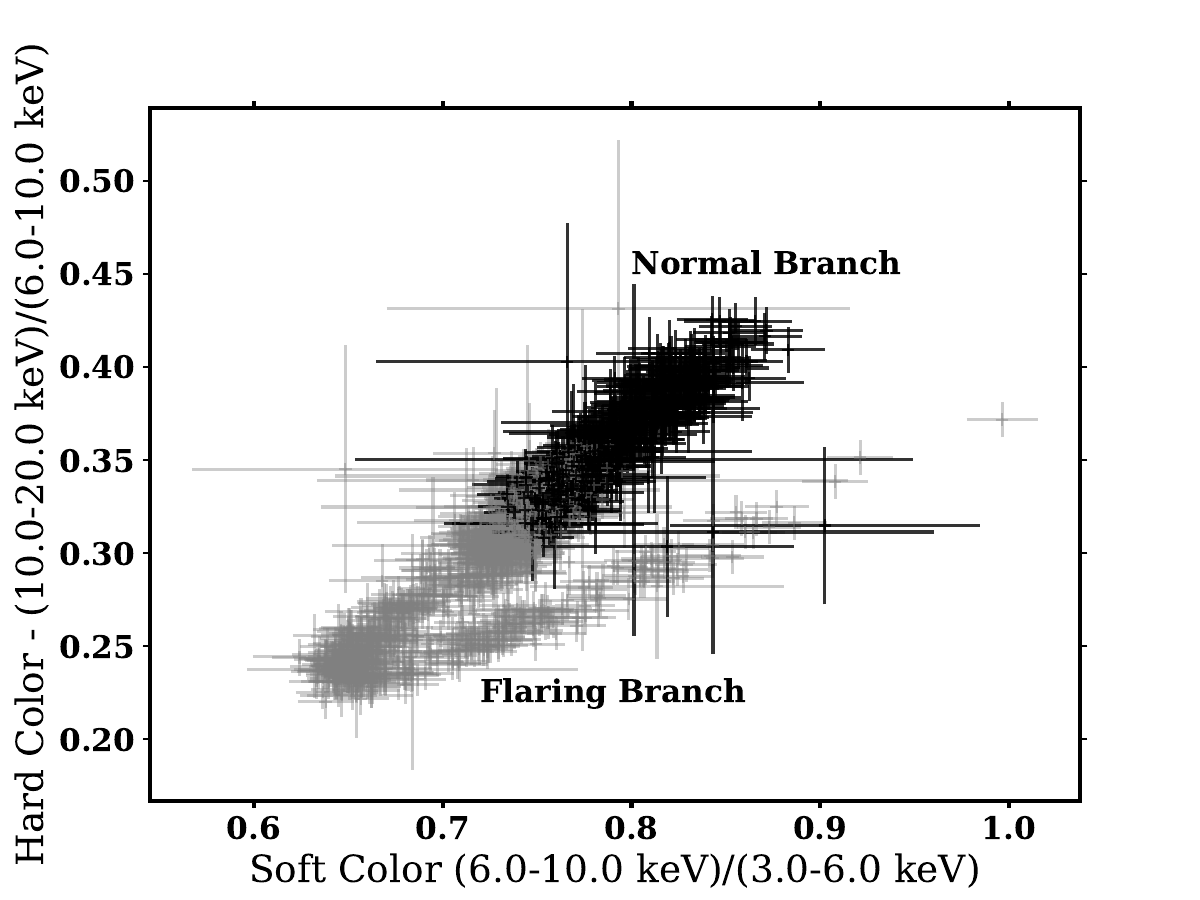}
\includegraphics[width=0.5\textwidth]{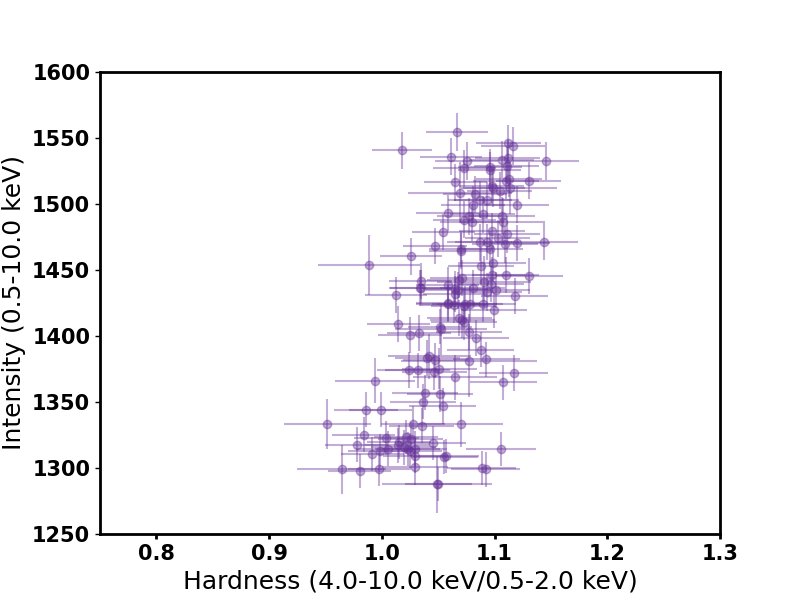}

\caption{Top: Color-color diagram constructed from all NuSTAR observations of GX~17$+$2 (ObsIDs starting with 30101023, 30902026, and 91001339). Time bins of 128 are used. The CCD indicates that the source was in the NB branch of its Z-track during our IXPE observations (black data points; gray are all previous NuSTAR observations of GX~17$+$2). Bottom: Hardness–intensity diagram constructed from the NICER observations simultaneous to our IXPE observation of GX~17$+$2. The NICER and NuSTAR CCDs indicate that the source was in the NB branch of its Z-track during the IXPE observations.}
\label{fig:hid}
\end{figure}

\begin{table*}
\centering

\caption{ IXPE, NICER, NuSTAR, and VLA  Observations of GX~17$+$2 (see Section \ref{sec:obs}). }

\begin{tabular}{c c c c c}
\hline
Instrument & Observation ID & Date (dd-mm-yyyy) & Start time (hh:mm:ss.ss) & Exposure time (ks)  \\ \hline

IXPE &03003501&23-10-2024--26-10-2024& 22:49:09.18   & 94.5 \\
NICER &7050410108-11& 22-10-2024--25-10-2024&22:19:40.00 & 1.1  \\
NuSTAR& 91001339002 & 23-10-2024-24-10-2024& 20:01:09 & 11.2 \\
& 91001339004 & 25-10-2024-26-10-2024 & 15:11:09 & 9.9 \\
VLA & 24A-387 & 08--08--2024 & 04:01:39.00 & 3.4 \\

\hline
\label{tab:table1}
\end{tabular}
\end{table*}

\begin{table}
\centering
\caption{Results obtained from the pcube analysis. The uncertainties mentioned are 1 $\sigma$ error (see Section \ref{sec:model_ind_analysis}).}

\begin{tabular}{c c c }
\hline
Energy band& PD (\%) & PA (\degr)\\ 
\hline
2-8 keV &$1.9\pm0.3$& $11\pm 4$ \\
2-4 keV & $2.0\pm0.3$& $9\pm3$\\
2-3 keV &$2.0\pm0.4$& $15\pm5$\\
3-4 keV &$2.0\pm0.4$& $3\pm5$\\
4-6 keV &$1.7\pm0.4$& $14\pm7$\\
6-8 keV &$2.2\pm1.0$& $10\pm13$\\
4-8 keV &$1.8\pm0.5$& $13\pm 7$\\
\hline
\label{tab:table2}
\end{tabular}
\end{table}

\subsection{IXPE}
\begin{table}
\centering
\caption{Best-fitting spectral model parameters from a multicolor disk blackbody component ({\tt diskbb}), thermal Comptonization component({\tt nthcomp}) and a reflection component ({\tt relxillNS}) model {\tt tbnew*(diskbb+nthcomp+relxillNS)*polconst*const} to the joint NICER, NuSTAR and IXPE spectra of GX~17$+$2. The uncertainties are 1$\sigma$. The calibration constant for NuSTAR FPMA is fixed at unity (see Section~\ref{sec:spec_pol_analysis}).}

\begin{tabular}{c c c c c c c c c}
\hline
Parameters& & & & & & \\
\hline
&tbabs&\\
$\text{N}_{\rm H}$ ($10^{22}$ atoms ${\rm\,cm}^{-2}$) & $3.29^{+0.03}_{-0.03}$&\\
$\text{N}_{\rm Si}$ ($10^{16}$ atoms ${\rm\,cm}^{-2}$)&$2.27^{+0.09}_{-0.10}$&\\
\hline
&diskbb&\\
$\text{kT}_{\rm in}$ (keV) &$1.15^{+0.03}_{-0.05}$&\\
DBB Norm&$303^{+19}_{-23}$&\\
\hline
&nthcomp&&\\
$\Gamma$& $2.10^{+0.02}_{-0.02}$&\\
$\text{kT}_{\rm e}$ (keV) &$3.53^{+0.02}_{-0.02}$&\\
$\text{kT}_{\rm BB} (keV) $&$1.51^{+0.03}_{-0.06}$&&\\
inp type&$0^{a}$&&\\
Redshift&$0^{a}$&&\\
Norm&$0.13^{+0.006}_{-0.006}$&\\
\hline
&relxillNS&&\\
Index1, Index2&$3^{a}$&\\
$\text{R}_{\rm br}$ ($GM/c^{2}$)&$990^{a}$&\\
a&$0^{a}$&\\
Incl&$30^{a}$&\\
$\text{R}_{\rm in}$ (in units of ISCO)&$7.9^{+1.5}_{-2.2}$&\\
$\text{R}_{\rm out}$ ($GM/c^{2}$)&$R_{\rm br}$&\\
z&0&\\
$\text{kT}_{\rm bb}$ &$\text{kT}_{\rm BB}$&\\
logxi&$2.90^{+0.10}_{-0.10}$&\\
Afe &$2^{a}$&\\
logN&$19^{a}$&\\
refl frac&$-1^{a}$&\\
norm ($\times 10^{-3}$)&$2.06^{+0.12}_{-0.12}$&\\
&Cross-calibration &\\
FPMB&$0.986^{+0.001}_{-0.001}$&\\
NICER&$0.926^{+0.001}_{-0.001}$&\\
DU1&$0.839^{+0.001}_{-0.001}$&\\
DU2&$0.839^{+0.001}_{-0.001}$&\\
DU3&$0.820^{+0.001}_{-0.001}$&\\
&Gain shift$*$&\\
\\

$\chi^{2}$/DOF&2801/2783& \\

\hline
$\text{Flux}^{b}$ ($10^{-9}$ ${\rm\,ergs/cm^{2}/s}$)&&\\
$\text{F}_{\rm Total}$&$13.08^{+0.04}_{-0.04}$&\\
$\text{F}_{\rm diskbb}$ &$5.85^{+0.19}_{-0.46}$&\\
$\text{F}_{\rm nthcomp}$ &$6.10^{+0.35}_{-0.19}$&\\
$\text{F}_{\rm relxillNS}$ &$1.12^{+0.08}_{-0.06}$&\\

\hline
\end{tabular}
\begin{flushleft}
\footnotesize{$^a$ fixed parameters}\\
\indent\footnotesize{$^{b}$ Energy flux of different model components in the 2-8~keV energy range } \\

\indent\footnotesize{*We apply gain fit using the slope of 1.02 (NICER) and 1.01-1.02 (IXPE DUs) and the gain shift of 48 keV (NICER) and 18-46 kev (IXPE DUs)}

\end{flushleft}

\label{tab:table3}
\end{table}

\begin{table*}
\centering

\caption{ PD and PA of each spectral component obtained considering the cases with fixed PA and PD of the {\tt diskbb} component and (1) Linked PA $\text{PA}_{\rm nthcomp}=\text{PA}_{\rm diskbb} \degr$ set-up, (2) Linked PA $\text{PA}_{\rm nthcomp}=\text{PA}_{\rm diskbb}-90 \degr$ set-up, (3) Linked PA $\text{PA}_{\rm nthcomp}=\text{PA}_{\rm diskbb} \degr$ set-up and the PA attributed to the reflection component being perpendicular to the disk PA, $\text{PA}_{\rm relxillNS}=\text{PA}_{\rm diskbb}-90 \degr$, from the best-fit spectro-polarimetric models {\tt tbnew*(diskbb+nthcomp+relxillNS)*polconst*const} and {\tt tbnew*(diskbb*polconst+nthcomp*polconst+relxillNS*polconst)*const} to the joint NICER, NuSTAR and IXPE spectra of GX~17$+$2. The uncertainties reported are  1$\sigma$ (see Section~\ref{sec:spec_pol_analysis})}
\begin{tabular}{ c c c c c c c c c c c c}\\
\hline
Set-up&Component& PD (\%)& PA (\degr)&$\chi^{2}$/DOF&\\
\hline
Linked PA \& Fixed disk polarization &diskbb&0.5&4&2802/2782&\\
&nthcomp&$<$3.4&$\text{PA}_{\rm diskbb}$&&\\
&relxillNS&$17.2^{+2.6}_{-13.8}$&$11^{+42}_{-4}$&&\\
\\
Linked PA \& Fixed disk polarization &diskbb&0.5&4&2802/2782&\\
&nthcomp&$<$6.6&$\text{PA}_{\rm diskbb}-90\degr$&&\\
&relxillNS&$26.4^{+24.8}_{-11.4}$&$8^{+6}_{-2}$&&\\
\\
Linked PA \& Fixed disk polarization &diskbb&0.5&4&2806/2783&\\
&nthcomp&$3.2^{+2.1}_{-0.4}$&$\text{PA}_{\rm diskbb}$&&\\
&relxillNS&$<$10.7&$\text{PA}_{\rm diskbb}-90\degr$&&\\
\\

\hline
Overall&&$1.6^{+0.2}_{-0.2}$&$10^{+3}_{-3}$&2801/2783\\
\\

\end{tabular}

\label{tab:table4}

\end{table*}

IXPE observed GX~17$+$2 from 2024 Oct 23, 22:49:09.184 UTC to Oct 26, 02:50:40.454 UTC (Obs ID: 03003501, PI: U. Kashyap) with a total live time of approximately 94.5 ks for each detector unit (DU) (see Table~\ref{tab:table1} and the light curve in Figure~\ref{fig:lc}). Spectral and polarimetric analysis was performed using HEASOFT version 6.33, with the IXPE Calibration Database (CALDB) version 20240125\footnote{\url{https://heasarc.gsfc.nasa.gov/docs/ixpe/caldb/}}. For extracting images and spectra, {\tt XSELECT}, available as a part of the {\tt HEASoft 6.33} package, was used extensively. Source photons were selected from a circular region with a radius of $60\arcsec$ for I, Q, and U spectra for each detector unit centered at the brightest pixel located at RA of 274$\fdg$00 and DEC of -14$\fdg$03. The weighted scheme  NEFF was adopted for the spectro-polarimetric analysis with improved data sensitivity\footnote{\url{https://heasarc.gsfc.nasa.gov/docs/ixpe/analysis/IXPE_quickstart.pdf}} \citep{2022SoftX..1901194B, 2022AJ....163..170D}.  The ancillary response files (ARFs)
and modulation response files (MRFs) were generated for each DU
using the {\tt ixpecalcarf} task, with the same extraction
radius used for the source region. GX~17$+$2 being a bright Z-source, following the prescription by \cite{2023AJ....165..143D}, we did not implement background rejection or subtraction. The unweighted model-independent polarimetric analysis was performed using the {\tt IXPEOBSSIM} package version 31.0.1 \citep{2022SoftX..1901194B}.

\subsection{NICER}
The Neutron star Interior Composition Explorer (NICER) observed GX~17$+$2 from 2024 Oct 22,  22:19:40.00 UTC to Oct 25, 21:37:20.00	UTC. The observation details are summarized in Table~\ref{tab:table1}. The NICER/X-ray Timing Instrument (XTI) observations were reduced using the NICER Data Analysis Software (NICERDAS) distributed with HEAsoft 6.33,
the CALDB 20240206\footnote{\url{https://heasarc.gsfc.nasa.gov/docs/heasarc/caldb/nicer/}}, and updated geomagnetic data. Cleaned event files were generated using the {\tt nicerl2} pipeline, applying standard filtering criteria, limiting undershoot rates (per focal plane module) to $<$ 500 cts/s and overshoot rates to $<$ 30 cts/s. We employed the \texttt{nicerl3-spect} task to generate source and background spectra using the {\tt SCORPEON} background model\footnote{\url{https://heasarc.gsfc.nasa.gov/docs/nicer/analysis_threads/scorpeon-overview/}} and the detector responses. We note that since the NICER observations carried out during orbit day observations are often affected by the optical light leak, we used orbit-night observations for the spectro-polarimetric analysis. These criteria resulted in a filtered exposure time of 1.1~ks.

\subsection{NuSTAR}

The Nuclear Spectroscopic Telescope Array (NuSTAR) is a hard X-ray focusing telescope, comprising two focal plane modules (FPMs), which operates over 3--79~keV \citep{2013ApJ...770..103H}. NuSTAR observed GX~17$+$2 for a total filtered exposure of $\sim$21.1~ks over two Target-of-Opportunity observations across 2024 October 23-26 (ObsIDs 91001339002 and 91001339004, PI: M. Ng). To reduce the data, we utilized the \texttt{nupipeline} tool from the NuSTAR Data Analysis Software (NuSTARDAS) v2.1.4 from HEASoft v6.34 (CALDB dated 20241104). We also used the \texttt{nuproducts} tool to extract energy-dependent light curves, and spectra, including branch-resolved products (see Figure~\ref{fig:hid}). Time bins of 128 s were used for the light curves, and the spectra were regrouped to have a minimum of 30 counts per bin. The scientific products were generated by extracting source events from a circular region with radius $120\arcsec$ centered at (R.A., Dec.) = ($274\fdg0071$, $-14\fdg0358$), along with a source-free background region nearby of the same size centered at (R.A., Dec.) = ($274\fdg0915$, $-14\fdg1074$).

\subsection{VLA}

GX~17+2 was observed with the VLA under Project Code 24A-387 (PI: E. C. Pattie) on 2024 Aug 08, with $\sim$1 hour on source at X-band (3-bit samplers at 8--12\,GHz with 4\,GHz of bandwidth). The data were obtained from the National Radio Astronomy Observatory (NRAO) data archive with a standard VLA pipeline calibration applied. The flux, bandpass, and polarization angle calibrator was 3C286 (J1331$+$030), the complex gain calibrator was J1832$-$1035, and the polarization leakage calibrator was J1407$+$2827. Data were processed in Common Astronomy Software Applications (CASA; \cite{2022PASP..134k4501C}), starting with the Science Ready Data Product, and calibrated for polarization using standard tasks. Data were imaged with \texttt{tclean} (\texttt{weighting=natural}) and phase self-calibrated (per scan, per spectral window). Polarization images were produced with \texttt{immath}.

\section{Results}
\label{sec:results}
\subsection{Source Spectral State}
Figure~\ref{fig:lc} shows the NICER and NuSTAR light curves of GX~17$+$2 during our IXPE observations. As Z-track patterns were not discernible in the IXPE HID or CCD, we used quasi-simultaneous NuSTAR and NICER observations to characterize the source spectral state during the IXPE observations.  The top panel of Figure~\ref{fig:hid} shows the NuSTAR CCD, where the hard and soft colors are defined as the ratios of the count rates in the energy bands between 10.0--20.0~keV and 6.0--10.0~keV, and between 6.0--10.0~keV and 3.0--6.0~keV, respectively. The NuSTAR CCD indicates that the source was detected in the NB during the two NuSTAR pointings. The bottom panel of Figure~\ref{fig:hid} shows the NICER HID, with the hardness defined as the ratio of the count rates in the energy bands between 4.0--10.0~keV and 0.5--2.0~keV. The NICER HID  shows a similar trend along the Z-track with no indication of source state transitions. To investigate the source behavior further during our observations, we performed detailed spectro-polarimetric investigations.

We note here that, in order to search for any possible state changes during the IXPE observations with no simultaneous NuSTAR and NICER observations, we also generated a 1 day-binned CCD considering the entire IXPE observation. We do not observe any strong variation in the spectral hardness, thus implying that GX~17$+$2 was entirely in the NB of its Z track throughout the IXPE observations.

\subsection{Model-independent polarimetric analysis}
\label{sec:model_ind_analysis}

\begin{figure*}
\centering
\includegraphics[width=0.45\textwidth]{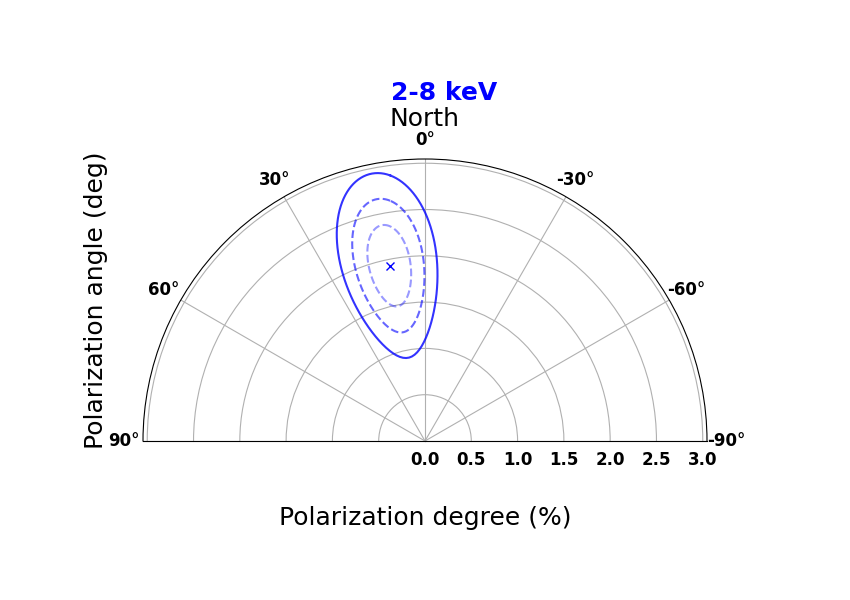}
\includegraphics[width=0.45\textwidth]{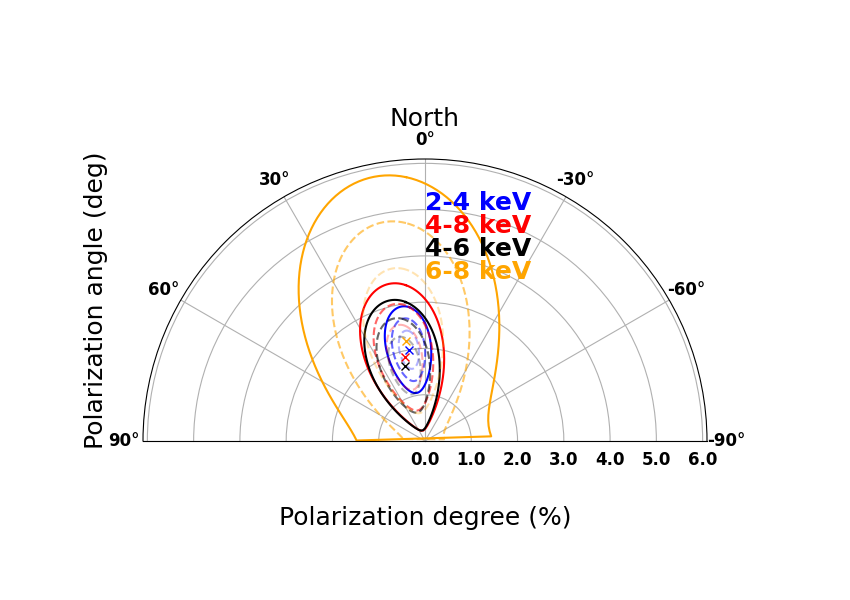}

\caption{Left panel: Contour plots of the polarization degree and angle, determined with the {\tt PCUBE} algorithm, at the 68 \%, 95 \% and 99 \% confidence levels, in the 2–8 energy band. Right panel: Contour plots of the polarization degree and angle, determined with the {\tt PCUBE} algorithm, at the 68 \%, 95 \% and 99 \% confidence levels, in the 2–4 keV (blue), 4-8 keV (red), 4-6 keV (black), and 6-8 keV (orange) energy bands. The contour plots show no strong variation of the source PA/PD with energy, considering the uncertainties.  }
\label{fig:mod_ind_pol}
\end{figure*}

\begin{figure}
\centering
\includegraphics[width=0.52\textwidth]{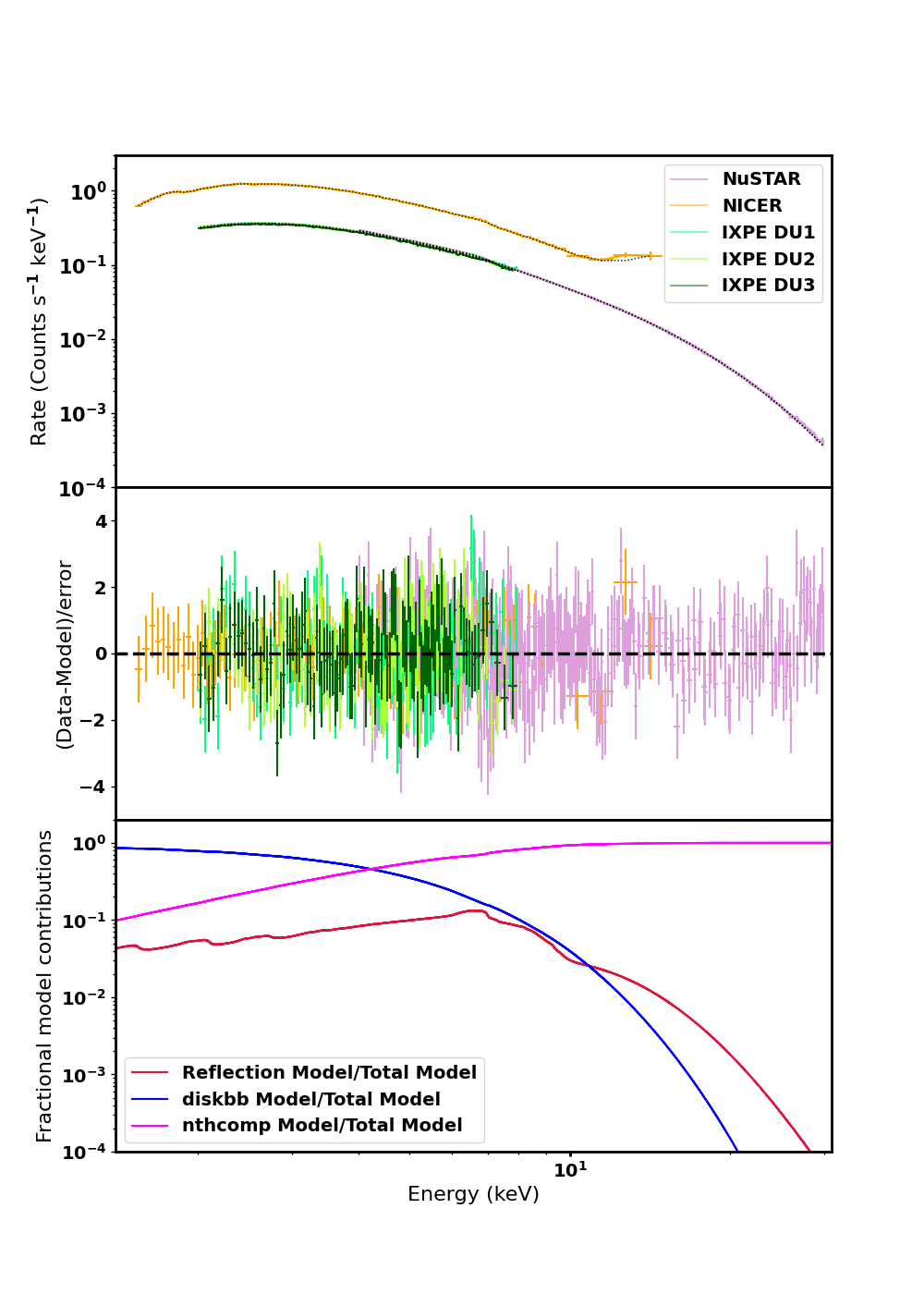}

\caption{First panel: Model fitted joint spectra of GX~17$+$2 as observed by IXPE DU1 (spring green), IXPE DU2 (green yellow), IXPE DU3 (dark green), NICER (orange) and NuSTAR (plum). The spectra are fitted with the {\tt tbnew*(diskbb+nthcomp+relxillNS)*polconst*const}
model in the 1.5--30~keV energy band. The total model is shown with the dotted black (NICER, NuSTAR, and IXPE). The spectral components are consistent with the previously reported X-ray studies of the source. Second panel: The residuals between the data and the best fit model.  Third Panel: The fraction of the different model components over the total model. The data for the IXPE, NICER, and NuSTAR spectra are rebinned only for plotting and representation purposes. }
\label{fig:spec_all}
\end{figure}

We carried out a polarimetric analysis of GX~17$+$2 using the {\tt ixpeobssim} package \citep{2022SoftX..1901194B} under the {\tt PCUBE} algorithm in the {\tt xpbin} tool. We employed the polarimetric analysis in the 2--8~keV, 2--4~keV, 4--6~keV, 6--8~keV, and 4--8~keV energy bands, applying unweighted analysis implemented in the {\tt ixpeobssim} package.
 The results obtained are reported in Table \ref{tab:table2} and Figure~\ref{fig:mod_ind_pol}, which shows a significant detection of polarization ${\rm PD} = 1.9 \pm 0.3\%$  with a polarization angle of ${\rm PA} = 11 \pm 4\degr$ from GX~17$+$2 in the 2--8~keV energy band.

\subsection{Spectro-polarimetric analysis}
\label{sec:spec_pol_analysis}

The spectral fitting and statistical analysis were carried out using the {\tt XSPEC v 12.14.0h} spectral fitting package distributed as a part of the {\tt HEASoft 6.33 package}. Considering the entire IXPE, NICER, and, NuSTAR observations, we fitted the joint spectra, with a model consisting of multicolor disk blackbody component ({\tt diskbb} in {\tt XSPEC}; \citet{1984PASJ...36..741M}), thermally Comptonized continuum component ({\tt nthcomp} in {\tt XSPEC}; \cite{1996MNRAS.283..193Z,1999MNRAS.309..561Z}), and a reflection component ({\tt relxillNS} in {\tt XSPEC}; \cite{2022ApJ...926...13G}) components. We modeled the
entire reflection spectrum using {\tt relxillNS} with a thermal input spectrum, $\text{kT}_{\rm BB}$, from the surface or BL of the NS. We also tied the inner and outer emissivity
indices to create a single emissivity profile, q, making the breaking radius $\text{R}_{\rm break}$
obsolete and fixing it at 3 \citep{2018MNRAS.475..748W}. The reflection fraction was fixed at -1 to obtain only the
spectral component due to the reflected fraction.  The
outer disk radius was set at 990 $\text{R}_{\rm g}$, inclination at 30\degr, and the spin parameter and iron abundance were fixed
at $a = 0$ and $\text{A}_{\rm fe} = 2$ \citep{2017ApJ...836..140L}, as otherwise, the parameters become unconstrained.  We performed the spectral fitting in the 2.0–-8.0~keV energy range (IXPE), 1.5--15.0~keV energy range (NICER), and 4.0--30.0~keV energy range (NuSTAR) to reduce the effect of background systematics in the spectrum. The total spectrum was modified by the presence of neutral hydrogen absorption in the interstellar medium, and this was accounted for by using the {\tt tbnew}\footnote{\url{https://pulsar.sternwarte.uni-erlangen.de/wilms/research/tbabs/}} model. The spectral fits showed a significant residual around $\sim$ 1.8~keV, which is likely the Si K edge, a known NICER instrumental systematic\footnote{\url{https://heasarc.gsfc.nasa.gov/docs/nicer/data_analysis/workshops/2024/joint2024.html}}. The Si abundance, when allowed to vary relative to the \texttt{wilms} abundance, was found to have a value of $2.27_{-0.10}^{+0.09}$ $\times 10^{16}$ atoms ${\rm\,cm}^{-2}$, which improved the fit significantly. The abundances and photoelectric cross-sections were adopted from \cite{2000ApJ...542..914W}. A constant ({\tt const}) model was used to account for the uncertainties in cross-calibration between NuSTAR FPMA, FPMB, NICER, and the IXPE DUs and is reported in Table~\ref{tab:table3}. We found significant residuals at low energies which is very likely due to the energy calibration uncertainties among instruments reported in the previous IXPE  \citep{2023A&A...676A..20U,2025ApJ...978L..19M} and joint NICER-NuSTAR studies \citep{2023ApJ...957...27M,2024A&A...688A.214B}. To account for this,  we applied a gain shift to the response files of NICER spectra and IXPE/I
spectra with the gain fit command in {\tt xspec} and linked the gain parameters of Q and U spectra to those of the I spectra for each IXPE DU. Figure~\ref{fig:spec_all} shows the NICER (orange), NuSTAR (plum), and IXPE (spring green, green
yellow, and dark green) spectra along with the best-fitting models.

 To study the polarization of the spectral components during our IXPE observations, we applied the {\tt polconst} model to the entire continuum model, {\tt tbabs*(diskbb+nthcomp+relxillNS)*polconst*const}, to check the consistency with the results obtained from the model-independent analysis.  The PD and the PA obtained from the spectro-polarimetric analysis are consistent with the results obtained for different energy bins from the model-independent analysis (see Section~\ref{sec:model_ind_analysis}). The corresponding best-fitting values obtained from the spectro-polarimetric analysis are reported in Tables~\ref{tab:table3} and \ref{tab:table4}.

Finally, we applied {\tt tbabs*(diskbb*polconst+
nthcomp*polconst+relxillNS*polconst)*const} model that assumes different constant polarization for each spectral component. However, only an upper limit of the PD of the {\tt nthcomp} component could be obtained and the PD of the {\tt relxillNS} component remains unconstrained.

 Due to the limited sensitivity of the data presented in this work and the limited energy coverage of IXPE, it is challenging to disentangle the polarization contributions from the individual components from the current spectro-polarimetric fitting. Thus, to investigate the polarization of the individual components and have better constraints on the polarization estimates, we considered two possible scenarios:  we estimated the polarization attributed to the Comptonization region and reflection component with the {\tt diskbb} PD fixed at the theoretically expected value ($\sim$ 0.5\%) for an inclination angle $\sim$ 30\degr \citep{1960ratr.book.....C,2022A&A...660A..25L} and PA fixed at 4 (obtained for the set-up with free PA of the individual components) with the the polarization attributed to the Comptonization region being parallel and perpendicular to the disk polarization; i.e., we allowed to vary only the PA of the {\tt nthcomp} component with the PA of {\tt diskbb} being linked by a relation $\text{PA}_{\rm nthcomp}$=$\text{PA}_{\rm diskbb}$ and $\text{PA}_{\rm nthcomp}$=$\text{PA}_{\rm diskbb}-90 \degr$. 

As the reflection is characterized by a polarization angle perpendicular to the disk plane \citep{1993MNRAS.260..663M,1996MNRAS.283..892P,2009ApJ...701.1175S}, we then consider scenarios where the reflection polarization is orthogonal to the disk polarization with the: the disk polarization being perpendicular and  parallel to that of the Comptonization region, with the fixed disk polarization. However, the former configuration yields unphysical results of polarization attributed to each component, so we report only the values obtained for the latter scenario. We report the results obtained from these spectro-polarimetric analysis in Table~\ref{tab:table4}.

We note here that we also consider a few cases in which different components are assumed to be unpolarized: (i) the disk and reflection components are unpolarized; (ii) the Comptonization and reflection components are unpolarized; and (iii) the disk and Comptonization components are unpolarized. While this approach does not necessarily yield a direct polarization estimate for the remaining component, we find that the polarization angle (PA) inferred for that third component (the disk, the Comptonization region, and the reflection component) remains the same as the overall PA obtained from the model-independent study and the spectro-polarimetric studies. We report these results in Table~\ref{tab:table6} in the appendix.

\subsection{The radio observation of GX~17$+$2}

\begin{figure}
\centering
\includegraphics[width=0.52\textwidth]{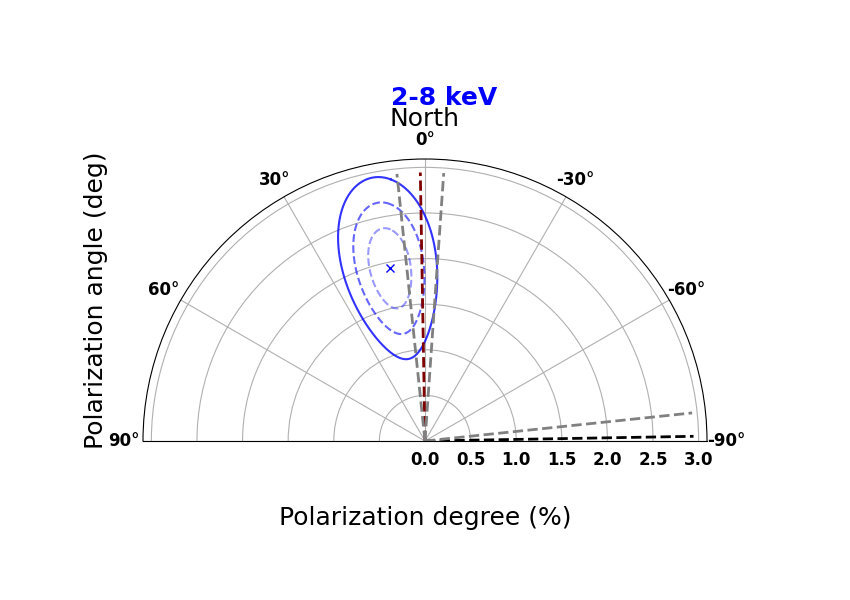}

\caption{A comparison of the X-ray PA and radio jet position angle obtained from our study of GX~17$+$2. The X-ray polarization degree and angle are determined with the {\tt PCUBE} algorithm, at the 68\%, 95\% and 99\% confidence levels, in the 2-8~keV energy band (in blue). The PA ($+1$\degr) and thus potential radio jet/jet axis position angles ($+1$\degr if parallel or $-89$\degr if perpendicular) are shown in maroon and black, with the gray lines representing the uncertainties on the radio jet position angle ($\pm5\degr$). The lower limit (-5\degr) on the uncertainty of the jet-axis position angle (-94\degr or 86\degr) goes beyond the plot. Our study shows the consistency between the X-ray PA and radio jet position angles in the case of GX~17$+$2.}
\label{fig:jet}
\end{figure}

GX~17+2 was strongly detected in radio as a point source at 10~GHz with an observation-averaged peak flux density of $5.139 \pm 0.005$~mJy/beam. We extracted radio polarization information from a continuous 17-minute light curve segment that displays stable flux density without strong variability, with an average flux density of $4.006 \pm 0.006$\,mJy and observed PA of $\sim12^\circ$ before Faraday correction.

We fitted for Faraday rotation \footnote{frequency-dependent rotation of the polarization angle as the radio wave travels through a magnetic field along the line of sight} \citep{Burn1966MNRAS.133...67B}) of the observed radio PA using two frequency subbands (8--10 and 10--12~GHz), following the standard calculation of $\mathrm{EVPA} = \mathrm{PA_{obs}} + \mathrm{RM}\,\lambda^{2}$, where $\mathrm{RM}$ is the rotation measure and $\lambda$ is the photon wavelength in meters. We find an RM of $\sim 210~\mathrm{rad~m^{-2}}$, corresponding to a 10\,GHz Faraday rotation angle of $+11^\circ$.

In the same 17-minute segment of the radio observation, we find a PD of $2.2\%\pm0.2\%$ and a Faraday de-rotated jet position angle (electric vector position angle; EVPA) of $1^\circ\pm5^\circ$. The spectral index ($\alpha$ defined as the flux density $S_\nu \propto \nu^\alpha$ at frequency $\nu$) of this time segment is $-0.78\pm0.02$. A more detailed radio analysis will be presented in E.C. Pattie et al., in preparation.

\section{Discussion}
\label{sec:discussion}

\begin{table*}
\centering
\scriptsize
\caption{The X-ray polarization properties of the Z-source NS LMXBs observed by IXPE during different spectral states, including polarization degree (PD), polarization angle (PA), and the X-ray PA with respect to the radio jet position angle detected from the sources (see Section~\ref{sec:discussion}).}
\begin{tabular}{ c c c c c c c}
\hline
Source  & PD  (\%) & State & PA (\degr) &PD/PA variation &X-ray PA w.r.t.  & ref \\ 
&(2--8 keV)& & & with energy&Radio jet&\\
\hline
&& & Sco-like Z-sources & &&\\

Sco~X-1  &$1.0\pm0.2$ &SA/FB& $8\pm6$& No&46\degr&\cite{2024ApJ...960L..11L}\\
GX~349$+$2  &$1.1\pm0.3$ &NB,SA,FB& $32\pm6$& No&-&\cite{2025arXiv250500813K}\\
GX~17$+$2  &$1.9\pm0.4$ &NB& $11\pm4$& No&PA aligned, jet aligned or orthogonal&This work\\

\hline
&& & Cyg-like Z-sources & &&\\
GX~5$-$1 &$3.7\pm0.4$& HB &$-9\pm3$ &PA&PA aligned, jet aligned or orthogonal$^{*}$&\cite{2024AA...684A.137F}\\
 &$1.8\pm0.4$ &NB/FB&$-9\pm6$ &PA&&\cite{2024AA...684A.137F}\\
 GX~340$+$0 &$4.02\pm 0.35$& HB &$37.6 \pm 2.5$&PA &&\cite{2024arXiv240519324B}\\
& $1.22\pm0.25$& NB & $38\pm6$ & PD  &  &\cite{2024arXiv241100350B} \\
Cyg~X-2  &$1.8\pm0.3$& NB&$140\pm4$&PD &aligned&\cite{2023MNRAS.519.3681F}\\
\hline
\label{tab:table5}
\end{tabular}
\begin{flushleft}
\footnotesize{$^*$ Pattie et al. in prep; the preliminary results obtained from the recent VLA observations shows intrinsic radio polarization angle of the radio jet of GX~5--1 is $\sim -18^{\circ}$, aligned in particular with the higher energy X-ray polarization angle. We note that in the cases of GX~17+2 here and GX~5--1 that the observed radio EVPA is aligned with the X-ray PA, and we assume that the jet is also most likely to be parallel with the radio EVPA, but cannot rule out the jet being perpendicular as well.}

\end{flushleft}

\end{table*}

In this paper, we report the first X-ray and radio polarization study of the NS Z-source GX~17$+$2 using IXPE and VLA.  Figure~\ref{fig:hid} shows the simultaneous NICER and NuSTAR observation of the source, indicating that the source was in the NB of its Z-track during our IXPE observations. The X-ray spectral studies show that the source spectrum is well described by a multitemperature disk blackbody component with accretion disk temperature  $\text{kT}_{\rm in}$=1.15~keV, along with a {\tt nthcomp} component representing the Comptonization of the seed photons
emitted by the NS surface in a hot plasma of electron temperature $\text{kT}_{\rm e} \sim$ 3.53 keV.  The reflection features observed in the joint spectra are well represented by the reflection
model {\tt RELXILLNS} with an assumption of a blackbody ($\text{kT}_{\rm BB}$ $\sim$1.51 keV) irradiating the disk. The X-ray spectrum is consistent with the previously reported spectra of the source \citep{2017ApJ...836..140L}, and recent studies of Sco-like Z-source \citep{2024ApJ...960L..11L}.

The model-independent polarimetric analysis shows a polarization of ${\rm PD} = 1.9 \pm 0.3\%$ with a ${\rm PA} = 11 \pm4 \degr$ in the 2--8~keV energy band, consistent with the results obtained from the spectro-polarimetric analysis.

While the polarization properties of the individual spectral components are poorly constrained with the IXPE data presented in this work, we discuss several test scenarios and examine their polarizations.

Assuming a disk polarization (${\rm PD}=0.5\%$) attributed to an electron scattering–dominated, optically thick accretion disk \citep{1960ratr.book.....C,2022A&A...660A..25L} for an inclination angle of $\sim30 \degr$, we obtained upper limits of the ${\rm PD} < 3.4\%$ and ${\rm PD} < 6.6\%$ attributed to the Comptonization component for the cases where the disk PA is assumed to be parallel and orthogonal to the PA of the Comptonization component. We also considered cases where the PA attributed to the reflection component is orthogonal to the disk, and PA of the disk aligned with the Comptonization component, which yields a PD of $3.2^{+2.1}_{-0.4}$ attributed to the Comptonization component. Using the relation for the optical depth for an optically thick Comptonization component reported in \citep{1996MNRAS.283..193Z}, we obtain the optical depth $\tau$ from the asymptotic power-law photon index $\Gamma\sim2.10$ of {\tt nthcomp} component as $\tau \sim5$. The estimated PD and upper limits on the PD attributed to the Comptonization region (obtained for different test scenarios) are compatible with the PD for a Thomson optical depth $\tau\sim5$ and $kT_{e}= 3.53{\rm\,keV}$ (see Figure 5 in \cite{1985A&A...143..374S}).

 In the case of SL, in WMNS, an overall low polarization is expected, along with a strong dependence of the PA on energy \citep{2024A&A...684A..62F,2025A&A...696A.181B}. Our joint spectral analysis shows a crossover between the two models representing the softer disk and the harder Comptonized emission (see Figure~\ref{fig:spec_all}). Thus, if the Comptonization component is associated with a spreading layer (SL), an energy dependence of the polarization angle (PA) is expected. However, the polarimetric results (particularly the model-independent results) reported in this work do not show any such dependence of PA on energy, and rather, the PA remains constant, making the SL scenario possibly unlikely in this case. Alternatively, the Comptonized component could also be a slab-like corona, along with a softer disk emission discussed in \cite{2022MNRAS.514.2561G}. A BL coplanar with the accretion disk could also account for the harder emission component, with the PA of the BL aligned with the PA of the softer disk emission.

Our study also demonstrates that, in every case (see Table~\ref{tab:table4}), the estimated PD attributed to the reflection component is consistent with the expected polarization estimations from radiation reflected by the accretion disk \citep{1993MNRAS.260..663M,1996MNRAS.283..892P}. Although the flux contributed by the disk and the Comptonization component changes over the IXPE energy band, the third panel of Figure~\ref{fig:spec_all} shows that the contribution of the reflection component remains constant (with the assumption of an unpolarized reflection line, which has some contribution to the flux variations). Additionally, the model-independent analysis shows no significant variation of the PA with energy (see Table~\ref{tab:table2}). This may suggest that the observed polarization and hence the observed PA from the source is possibly dominated by the polarization attributed to the reflection component (however, see discussions in section~\ref{jet}).

We note here that the data statistics presented in this work do not allow us to constrain the variation in the PA. 
Moreover, given the limitations of the data presented in this work -- with IXPE's narrow (2--8~keV) energy coverage and limited sensitivity, it is essentially not possible to disentangle the spectral components and the polarization associated with the individual components. Hence, we refrain from further interpretations of the source geometry with the results presented in this work. However, future observations with polarimeters offering broad energy coverage such as XPoSat\footnote{\url{https://www.isro.gov.in/XPoSat_X-Ray_Polarimetry_Mission.html}} \citep{2025ExA....59...17S}, the upcoming enhanced X-ray Timing and Polarimetry mission \citep{2025arXiv250608101Z}.  and proposed missions like REDSoX \citep{2017SPIE10397E..0KM} or PolStar \citep{2016APh....75....8K} along with long IXPE exposures would help put better constraints on the source geometry.

 \subsection{Energy-dependent polarization properties of the Cyg-like and Sco-like Z-sources}
 
In Table~\ref{tab:table5}, we report the polarization estimation of both Cyg-like and Sco-like Z-sources obtained from the previous IXPE polarization studies. The estimated total polarization (${\rm PD} = 1.9 \pm 0.3\%$) of GX~17$+$2, is consistent with the polarization properties previously reported for the other two Sco-like sources: Sco~X-1 and GX~349$+$2 (see Table~\ref{tab:table5}). 

Recent studies show that, unlike the Cyg-like sources, the Sco-like sources exhibit no strong indications of the energy dependence of PD/PA (see Table~\ref{tab:table5}). 
As reported in Table~\ref{tab:table2} and Figure~\ref{fig:mod_ind_pol}, the energy-dependent analysis, in the case of GX~17$+$2, shows no evidence of polarization variations with energy -- which is also reported for Sco~X-1 and GX~349$+$2. This may suggest that the variations in polarization properties with energy may be intrinsic to the Sco-like sources indicating a fundamental difference between the Sco-like and Cyg-like Z-sources, possibly associated with the accretion geometry or mechanism of each class. We note that, given the limited number of sources in each class (Cyg-like and Sco-like), with limited sampling in different Z states (NB, HB, and FB), it is not possible to draw strong conclusions at present. Future observations will be crucial for enabling a more detailed investigation into the energy dependence of polarization properties in Cyg-like and Sco-like sources.

 \subsection{X-ray PA alignment with the radio-jet position angle of GX~17$+$2 }
\label{jet}
 Our study reports a detection of GX~17$+$2 in radio as a point source at 10~GHz with an observation-averaged peak flux density of $5.139\pm 0.005$~mJy with a PD of $2.2\pm0.2\%$ and an EVPA (Faraday de-rotated radio PA) of $1^\circ\pm5^\circ$. As shown in Figure~\ref{fig:jet}, a comparison of the X-ray and radio polarization shows that the estimated X-ray PA of GX~17$+$2 in the NB state of the Z-track is consistent with the radio EVPA (within uncertainties). 

Previous studies of BH LMXB jets found that the radio EVPA of the compact jet was aligned with the jet axis, as confirmed by resolved imaging \citep[e.g.,][]{Russell2015MNRAS.450.1745R, Rushton2017MNRAS.468.2788R}. We thus also assume this is the most likely interpretation for the jets of the NS LMXBs, though we cannot rule out that the jet may be perpendicular to the observed radio EVPA; the orientation of the jet relative to the radio EVPA depends on the configuration of the jet's magnetic field, which is unknown here. Thus, under this assumption, the X-ray PA estimated in the case of GX 17+2 is parallel to the radio jet position angle, which is consistent with the previous studies that reported the alignment of the X-ray PA (especially in the hard energy) with the radio-jet position angle for a few Z-sources: Cyg~X-2 \citep{2023MNRAS.519.3681F} and GX~5$-$1 (E.C. pattie et al, in preparation). These studies also favor an SL at the NS surface as the potential source of the polarization signal \citep{2023MNRAS.519.3681F}. 

However, at this stage, we cannot rule out the possibility that the estimated X-ray position angle is orthogonal to the radio jet axis (see Figure~\ref{fig:jet}). An orthogonal radio PA may suggest that the X-ray polarization observed in the case of GX 17+2 might be associated with the direct emission from the accretion disk or a BL, which is coplanar with the disk. Moreover, we can not eliminate a scenario where the Comptonization region is vertically extended, with the PA orthogonal to the jet axis. So far, X-ray lags have been investigated for a sample of X-ray binaries, revealing changes in the lag by some factor as the source transitions between states \citep{2022ApJ...930...18W}.  If these lags are associated with the light travel time delays, the distance between the disk and the Comptonized component must vary, suggesting a vertically extended  Comptonization region \citep{2021A&A...654A..14D, 2022ApJ...930...18W}. Moreover, previous timing studies report some evidence of a changing size of the Comptonized region in the case of GX~17$+$2; however, more sensitive observations are required to validate these findings \citep{2019ApJS..244....5S}. We also note here that, if the jet axis is orthogonal to the observed X-ray PA, then, considering the fact that the reflected emission is polarized perpendicular to the disk plane \citep{1993MNRAS.260..663M,1996MNRAS.283..892P,2009ApJ...701.1175S}, the observed X-ray PA from GX~17+2 is less likely to be dominated by the reflected emission as this would indicate that the jet is aligned along the plane of the accretion disk, contrary to the theory of how jets are launched perpendicular to the inner disk \citep{1982MNRAS.199..883B}.

The previously studied Sco-like source, Sco~X-1, observed in the SA/FB state, shows a total polarization of $1.0 \pm 0.2\%$ \citep{2024ApJ...960L..11L} with a rotation of the X-ray PA by $46\degr$ with respect to the known radio jet position angle of the source \citep{Fomalont2001ApJ...558..283F}. 
As stated before, in the most typical scenario, the X-ray PA is expected to be either aligned or perpendicular to the radio jet position angle. Thus, the observed alignment or the difference (90\degr) with the radio jet position angle in the case of GX~17+2 invalidates the possibilities of X-ray PA and radio-jet position angle misalignment (at an angle that is neither perpendicular nor parallel, e.g., $45$\degr) being intrinsic to the Sco-like sources as detected in the case of Sco~X-1. Furthermore, the variations of polarization properties along the Z track have been reported for multiple Z-sources (see Table~\ref{tab:table5}). A hint of PA rotation in the FB  with respect to the SA/NB has also been reported in our previous studies in the case of Sco-like Z source GX~349$+$2 (see discussion of \cite{2025arXiv250500813K}).  Thus, there are possibilities of Sco~X-1 exhibiting PA swing during the FB with respect to the other Z states. Although the alignment of the X-ray PA and radio jet position angle has not yet been investigated for all the Sco-like Z-sources, our future VLA studies of GX~349$+$2, and possibly a confirmation of the jet axis using VLBI for GX~17+2 and GX~5--1, for which we have an accepted VLBA proposal, and future observations with further sampling of the different Z-source states will test the proposed hypothesis.

\section{Summary}
\label{sec:summary}

 This work reports the first-ever X-ray and radio polarization study of the Sco-like Z source GX~17$+$2. The X-ray source was discovered with a ${\rm PD} = 1.9 \pm 0.3\%$  at a polarization angle of ${\rm PA} = 11 \pm4 \degr$. The detailed spectro-polarimetric study shows the presence of a softer accretion disk emission, along with a harder Comptonized component. The NuSTAR spectra reveal the presence of reflection features consistent with previous studies of the source. Most interestingly, we report the VLA detection of the source at 10~GHz, with an estimated linear polarization of ${\rm PD} = 2.2 \pm 0.2\%$ and an EVPA of $1\pm5 \degr$, indicative of a radio-jet position angle most likely parallel (but we cannot rule out the perpendicular case) to the X-ray PA of GX~17$+$2. Our study manifests the requirement of future IXPE observations of the NS Z-sources tracing out the entire Z-track with longer exposures,  complemented by a broadband study using future X-ray polarimeters.

\section{Acknowledgments}
\begin{acknowledgments}
This research used data provided by the Imaging X-ray Polarimetry Explorer (IXPE), NICER (Neutron star Interior Composition Explorer), Nuclear Spectroscopic Telescope Array (NuSTAR) and distributed with additional software tools by the High-Energy Astrophysics Science Archive Research Center (HEASARC), at NASA Goddard Space Flight Center (GSFC). U.K. and T.J.M. acknowledge support by NASA grant 80NSSC24K1747. M.N. is a Fonds de Recherche du Quebec – Nature et Technologies (FRQNT) postdoctoral fellow.  A.J.T and P.B.C acknowledge that this research was undertaken thanks to funding from the Canada Research Chairs Program and the support of the Natural Sciences and Engineering Research Council of Canada (NSERC; funding reference number RGPIN--2024--04458).
We acknowledge Edward Nathan for the useful discussions on reflection modeling. We would like to thank Fabio La Monaca and Alessandro Di Marco for coordinating the NuSTAR observations simultaneously with our IXPE observations
and for notifying us that they had done so to increase the value of the IXPE observations for the community, which significantly helped our spectropolarimetric analysis.
\end{acknowledgments}


%

\vspace{5mm}
\facilities{IXPE, NICER, NuSTAR, VLA}


\software{ixpeobssim, xspec
, HEASoft
          }


\appendix

\begin{table*}
\centering

\caption{ PD and PA of each spectral component obtained considering the cases where we consider (1) unpolarized disk and reflection component (2) unpolarized comptonization and reflection component, and (3) unpolarized disk and comptonization component from the best-fit spectro-polarimetric model  {\tt tbnew*(diskbb*polconst+nthcomp*polconst+relxillNS*polconst)*const} to the joint NICER, NuSTAR and IXPE spectra of GX~17$+$2. The uncertainties reported are  1$\sigma$ (see Section~\ref{sec:spec_pol_analysis})}
\begin{tabular}{ c c c c c c c c c c c c}\\
\hline
Set-up&Component& PD (\%)& PA (\degr)&$\chi^{2}$/DOF&\\
\hline
Unpolarized disk \& reflection&diskbb&0&0&2806/2783&\\
&nthcomp&$3.8^{+0.5}_{-0.5}$&$9^{+3}_{-3}$&&\\
&relxillNS&0&0&&\\
\\
Unpolarized Compt \& reflection &diskbb&$2.8^{+0.4}_{-0.4}$&$10^{+3}_{-3}$&2808/2783&\\
&nthcomp&0&0&&\\
&relxillNS&0&0&&\\
\\
Unpolarized disk \& Compt &diskbb&0&0&2804/2783&\\
&nthcomp&0&0&&\\
&relxillNS&$20.1^{+2.5}_{-2.5}$&$9^{+3}_{-3}$&&\\

\hline
\\

\end{tabular}

\label{tab:table6}

\end{table*}

\pagebreak


\bibliography{sample631}{}
\bibliographystyle{aasjournal}



\end{document}